\documentclass[reviewmode]{AEA}

\usepackage{tabularray}
\usepackage{float}
\usepackage{graphicx}
\usepackage{codehigh}
\usepackage[normalem]{ulem}
\UseTblrLibrary{booktabs}
\UseTblrLibrary{siunitx}
\usepackage{subcaption}
\usepackage{placeins}
\usepackage{tikz}
\usepackage{pdflscape}

\NewTableCommand{\tinytableDefineColor}[3]{\definecolor{#1}{#2}{#3}}

\usepackage{scalerel,stackengine}
\stackMath
\newcommand\reallywidehat[1]{%
\savestack{\tmpbox}{\stretchto{%
  \scaleto{%
    \scalerel*[\widthof{\ensuremath{#1}}]{\kern-.6pt\bigwedge\kern-.6pt}%
    {\rule[-\textheight/2]{1ex}{\textheight}}
  }{\textheight}%
}{0.5ex}}%
\stackon[1pt]{#1}{\tmpbox}%
}
\usepackage{natbib}

\draftSpacing{1.5}

\begin{document}

\title{Have Data Centers Raised Your Electric Bill? \\Causal Evidence from the United States}

\shortTitle{Data Centers}
\author{Asa Watten, John Bistline, and Geoffrey Blanford
\thanks{Watten: EPRI (awatten@epri.com); Bistline: Watershed; Blanford: EPRI}}
\date{\today}

\begin{abstract}
We estimate that data centers caused average retail electricity rates to fall modestly in the United States from 2015 to 2024 using an instrumental variables approach. Despite prevailing sentiment, the finding is consistent with economic reasoning: existing large power system fixed costs, economies of scale in transmission and distribution, and declining unit costs for generation imply that durable demand growth lowers average prices. We find patterns of economies of scale for transmission, distribution, and generation costs as well as within and across retail customer classes. We caution that future supply constraints could reverse the effect.  
\end{abstract}

\maketitle

In 2024, data centers used about 4.5\% of electricity in the United States, more than double their consumption five years earlier, and are projected to account for 9-17\% by 2030 \citep{epri2026poweringintelligence}. Driven by a rapid acceleration in artificial intelligence and other data-intensive services, this growth has provoked backlash from some communities and their representatives in government. State legislatures have introduced bans on new data center construction, and numerous cities and towns have enacted moratoriums or resisted projects. Concerns about data centers include air pollution, noise, water consumption, and increasingly, higher utility bills. Perceived rate impacts are a central point of contention, motivating the March 2026 ``Ratepayer Protection Pledge Proclamation'' issued by the White House and signed by seven major technology companies. Roughly 70\% of Americans say they oppose data centers being sited near them \citep{gallup_poll2026}.

While ratepayers fear new load will raise prices, the historical record is consistent with the opposite. In the second half of the 20th century, US electricity demand increased while real prices decreased. In this century, both price and demand trends flattened. From 2014 to 2020 demand fell by 0.3\% year over year while average prices grew at a rate of 0.2\%, well below inflation during this period. From 2021 to 2024 demand grew on average a modest 1.5\% year over year while electricity rates grew at 5.2\%, in line with the average annual increase in core consumer prices of 5.8\%  \citep{eia2025epa}, meaning inflation adjusted electricity rates were flat on average. These recent rate increases have not been uniform across states but concentrated in the Northeast and West Coast \citep{wiser2025factors}. California in particular saw average prices increase nearly 40\% during the same period owing largely to wildfire costs \citep{SINGH2025107475}. States in which load increased tended to have below average price increases \citep{wiser2025factors}.

Observing the capacity expansion of data centers in particular, there does not appear to be a correlation with rising average prices at the state level from 2019 to 2024 (see Figure \ref{fig:price_vs_dc}). Virginia, which is the U.S. leader in data centers (where they consume over 20\% of electricity), showed electricity rate increases similar to the average. However, these correlations do not establish causality due to omitted variable bias from endogenous selection into states, which could flow in either direction. In this paper, we estimate the price changes caused by data centers over this period by deploying instrumental variables to isolate a plausibly exogenous source of variation, thus overcoming unobserved confounding drivers of price changes and data center construction. For the instrumental variable, we use the total length of the 1947 Eisenhower Interstate Highway Plan within each state, which we argue is exogenous to electricity prices conditional on GDP and population.

It seems plausible a priori that data center developers would be incentivized to locate in places where future electricity prices are expected to be low even with load growth, which would cause negative bias in the ordinary least squares (OLS) estimate. Alternatively, we can tell a speculative story that leads to bias upwards: utilities that clear interconnection queues faster have higher capacity to update their infrastructure; having more capacity to build is costly; and data centers prefer to locate in places where there is quick access to power, which may be correlated with higher prices. These are hypotheses; there may be other plausible stories. An IV approach is required to adjudicate between hypotheses.

We estimate for every 10\% increase in data center capacity average residential retail prices fell by approximately 0.4\% on average (Table \ref{tab:main_results}). The Anderson-Rubin test---which is robust to weak instruments (F = 7.8 in the preferred specification)---is significant at the 0.1\% level, clearly rejecting positive effects. The result is interpreted thus: from 2019 to 2024 the average residential customer lived in a state where data center capacity grew 160\%,\footnote{1.6 is the mean difference of the inverse hyperbolic sine of state-level data center capacity, which allows for zeros, weighted by the number of customers in each state. Our data show that overall data center capacity increased 3.5 times in this period. We exclude cryptocurrency mining and hosting.} which caused their rates to fall by 6\%. We present results for industrial and commercial average prices with smaller effect sizes in the Supplemental Appendix. We also estimate spillover effects from neighboring states and find an elasticity of 0.007; although not statistically significant and the instrument is quite weak. 

How can higher demand lead to lower prices? Does the law of supply and demand not apply? A key difference between the power sector and other commodity markets is that due to the natural monopoly of distribution networks and high fixed costs throughout the system, retail prices are a function, at least in part, of average costs and not marginal costs. If the incremental cost of meeting demand is smaller than the status quo mix of average costs and marginal costs that are passed through to customers, then new demand mechanically lowers average costs.\footnote{A version of this argument is presented in \cite{epri2025winwin}.}

Consider first a setting in which the physical assets of the power system are held fixed. As demand rises, grid operators dispatch electricity from increasingly expensive generators, which increases wholesale electricity prices as the law of supply and demand suggests. This is what happens when, for example, a greater number of air conditioners run as temperature rises on a hot day. These short-run variable costs are passed through to retail customers via wholesale energy prices or similar mechanisms, but unlike in most commodity markets, retail tariffs must also recover the fixed costs of generation, transmission, and distribution capital. A bump in demand can allow for higher utilization of the existing generation \citep{retancourt1987economies} and distribution network \citep{kwoka2005electric,yatchew2000scale}. Provided the bump does not violate capacity constraints, this leads to fixed costs that are spread across more kilowatt-hours (kWh), working in the opposite direction as the dispatch effect and pushing average costs down. 

If, instead, we allow the power system to change, and if higher demand is not within the expected distribution of demand but a new long-lived demand shift---such as new data centers, electric cars, and heat pumps---then the system will invest in new capital equipment to service this new load, shifting the dispatch curve outward and lowering marginal variable cost. Due to secular technology cost declines and efficiency gains, the incremental mix of new generation assets tends to have a lower average cost than the embedded incumbent asset mix \citep{bolinger2022levelized,way2022empirically,mirletz2024annual,cole2025cost,seel2025us}, thus lowering the amortized capital cost of generation per kWh.

We formalize a simple analytic model in the Supplemental Appendix. We find that as long as there are no constraints on building new capacity, a shift in demand will cause average prices to fall if the sum of the current marginal variable cost of electricity and the shadow cost of the resource adequacy constraint\footnote{This is equal to the levelized cost of capital for capacity net of the savings provided by shifting the dispatch curve outwards or the market clearing ``capacity lambda''.} is smaller than present average costs. We expect this to have been true nearly everywhere given observed average prices, marginal variable costs, and reasonable estimates of the levelized cost of capital for generating units at the margin (see the Supplemental Appendix for a numeric example). 

In addition to causal evidence, we present suggestive evidence for a mechanism: economies of scale across all categories of power system cost. Section \ref{sec:desc} presents descriptive statistics showing that the four largest sub-categories of power systems costs---capital expenditures (capex) and operational expenditures (opex) of both transmission and distribution (T\&D) and generation---fall per unit electricity delivered as demand increases on average. We also observe evidence of own-customer-class economies of scale and lower prices when other customer classes increase their demand, rejecting the notion of widespread cross-class subsidization of system costs. Section \ref{sec:causal} presents the IV analysis of data center effects.

This study adds to the literature on economies of scale in electricity markets since \cite{christensen1976economies}, and contributes to the recent literature on the determinants of retail electricity prices in the US (e.g., \citealt{wiser2025factors,borenstein2022two}) and on the potential rate effects of data centers (e.g., \citealt{knittel2025flexible}). It is unique in showing the causal effects of long-lived changes in electricity demand. While \cite{wiser2025factors} shows negative correlation between average prices and changes in electricity sales, the study does not identify causal effects due to the likely presence of selection bias. Our results underline the importance of the ``build margin'' when estimating the effect of long-lived demand. Existing studies that exploit short-run random variation are only able to capture the dispatch effect (``operating margin'') and exclude the effects of new generation, transmission, and distribution capacity that are driven by shifts in future demand expectations. For example, \cite{dong_etal_2025} use random variation in annual temperature to estimate the effect of demand changes on electricity prices, while \cite{holland2016there} use random variation in hourly demand by controlling for hour-month fixed effects to estimate marginal changes in local air pollution. This relates to an on-going debate over how to compute marginal emissions for energy demanding or supplying investments (see \cite{bistline2025emissions}).

\section{Data}\label{sec:data}

\subsection{Data Used for Utility-Level Analysis}

To investigate power system costs, we use data collected by the Federal Energy Regulatory Commission (FERC) Form 1 for CONUS. These data contain utility level data of costs broken down by T\&D and generation. These are further disaggregated into operation and capital expenses. The utilities represented in these data are exclusively large investor-owned utilities. We exclude utilities with any years with missing or no residential, industrial, or commercial customers, which removes wires-only utilities. Total sales of the remaining 366 utilities in the sample range from 3 gigawatt-hours to 162 terawatt-hours. These data account for 69\% of retail electricity sales. 

Importantly, FERC exempts some wholesale generators from reporting costs. Therefore, some generation costs are missing or show up exclusively in the generation opex of the utility buying the wholesale power. The extent to which this is an issue is discussed in Section \ref{sec:desc_utils}.

\subsection{Data Used for State-Level and Causal Analysis}

Retail revenue and sales come from US Energy Information Administration (EIA) Form 861 as cleaned and compiled by \cite{pudl_2026}. We use data from bundled (energy and wires) service providers only, which excludes behind the meter as well as wire and energy only providers. This accounts for 93\% of residential sales during this period. Generator capacity additions for solar, wind, natural gas, and energy storage are from EIA Form 861.

To measure data center expansion, we use data from \cite{SP_dckb}, which are compiled from various sources including public permitting records, site visits, briefings, and surveys. The data include the year in which data centers became operational or were expanded, along with nominal information technology capacity of each facility. We omit data centers for mining or hosting cryptocurrency because they have potentially very different load patterns. Capacity is missing for a non-trivial share of pre-2015 facilities; we therefore restrict our main analysis to additions between 2015 and 2024, a period that also coincides with the rapid expansion of hyperscale operators. Over these 10 years, these data show between 1.1 and 6.2 gigawatts of capacity were added annually to CONUS; 21.3 gigawatts in total.

Annual state-level population estimates are from  \cite{census_popest_2010s, census_popest_2020s}, while GDP estimates are from \cite{bea_gdp_state}. 

The 1947 Dwight D. Eisenhower National System of Interstate and Defense Highways Plan geographic data were downloaded from the National Bureau of Economic Research website, as used in \cite{brinkman2024freeway}.

\section{Descriptive Evidence of Economies of Scale}\label{sec:desc}

\subsection{Descriptive Evidence of Economies of Scale by Cost Type}\label{sec:desc_utils}

\begin{figure}[t!]
    \includegraphics[width=1\linewidth]{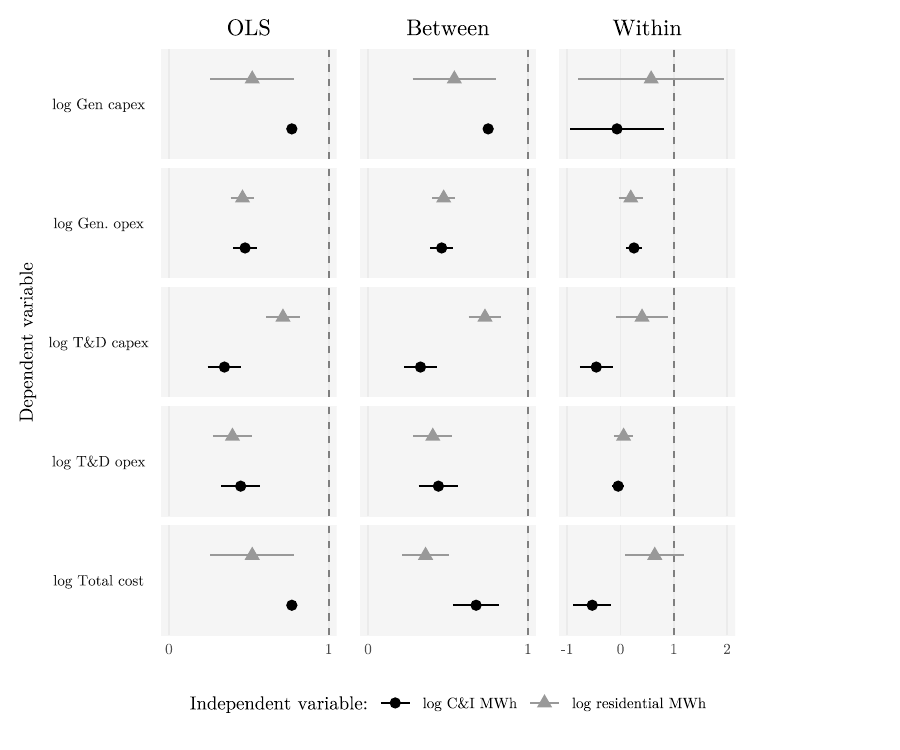}
\caption{Sectoral Demand and Costs} \label{fig:price_vs_dc}\label{fig:cost}
\begin{figurenotes}
The dependent variables are listed as rows and model specification is listed as columns with each combination showing estimates from a distinct regression. 
Column 1 is the ordinary least squares estimate, column 2 adds year fixed effects, and column three includes both year and utility fixed effects. The 95\% confidence interval are shown below the point estimate in brackets. All estimates are less than 1, indicating a negative correlation with average costs. There are between 414 and 440 observations in each model depending on missing data. The sample period spans 2015-2024.
\end{figurenotes}
\end{figure}

 In this subsection we explore the correlation between costs and sectoral demand and observe evidence that the correlation is concave for total costs and all cost categories individually. Using our panel of large investor-owned utilities, the four cost categories we examine are: operational expense (variable cost) and  capital expense (fixed cost) for generation and T\&D. Using a log-log specification allows us to estimate the relative rate of change of demand associated with a change in sectoral-costs. Thus, we estimate the equation:
\begin{equation}
    \ln c^j_{it}=\beta_1 \ln D^{\text{res}}_{it} + \beta_2\ln D^{\text{C\&I}}_{it} +\alpha_i + \delta_t + \varepsilon_{it}
\end{equation}
where $c^j_{it}$ is cost of type $j$, for utility $i$ in year $t$;  $D^{\text{res}}_{it}$ and  $D^{\text{C\&I}}_{it}$ are residential and the T\&D demand respectively; $\alpha_i$ and $\delta_t$ are utility fixed effects and year fixed effects. 

Figure \ref{fig:cost} displays the coefficient estimates for $\beta_1$ and $\beta_2$. The leftmost panel includes no fixed effects, the middle panel adds year fixed effects, and the last panel includes both year and utility fixed effects. Coefficients less than 1 indicate a concave correlation between cost and demand. Overall, the data show concave relationships, a pattern consistent with economies of scale and also reverse-causality.

Comparing between utilities, costs for utilities with higher demand increase at a lower rate than demand for both residential and C\&I customers. Looking at the last panel, as demand increase within a utility, costs are concave, and also sometimes negatively correlated with demand. As there is no theoretic reason we can think of for demand to cause a decrease in total costs, these negative coefficients on T\&D capital expense and total cost indicate that firms either run their equipment less often, increase the energy efficiency of their operation, or leave the utility service territory in response to higher prices. 

The literature estimates that both customer classes are price inelastic in the short-run with central estimates of about -0.1 \cite{burke2018price}. Adjusting estimates by adding 0.1 all coefficients remain below unity and thus concave.

\subsection{Descriptive Evidence of Economies of Scale by Customer Class}

\begin{table}
\caption{Cross-Sector Price and Demand Correlation}
\label{tab:demand_rev_corr}
\centering
\resizebox{1\linewidth}{!}{
\begin{talltblr}[         
caption = {},
  entry = none,
  label = none,
note{}={+ p < 0.1, * p < 0.05, ** p < 0.01, *** p < 0.001},
note{ }={The dependent variable is given by column-group title. Observations 
            are state-years. ``Between'' estimates make comparisons across states while 
            controlling for common shocks and trends; ``Within'' uses the variation 
            within states by adding state fixed-effects. ``Spills'' adds variables
            for adjoining state neighbors. The sample period spans 2014-2024.},
]                     
{                     
colspec={Q[]Q[]Q[]Q[]Q[]Q[]Q[]},
hline{2}={3,6-7}{solid, black, 0.03em},
hline{2}={2,5}{solid, black, 0.03em, l=-0.5},
hline{2}={4}{solid, black, 0.03em, r=-0.5},
hline{3}={1-7}{solid, black, 0.05em},
hline{11}={1-7}{solid, black, 0.05em},
hline{1}={1-7}{solid, black, 0.08em},
hline{15}={1-7}{solid, black, 0.08em},
column{3-4,6-7}={}{font=\fontsize{0.8em}{1.1em}\selectfont, halign=c},
cell{1}{1}={}{font=\fontsize{0.8em}{1.1em}\selectfont, halign=c},
cell{1}{2}={c=3}{font=\fontsize{0.8em}{1.1em}\selectfont, halign=c},
cell{1}{5}={c=3}{font=\fontsize{0.8em}{1.1em}\selectfont, halign=c},
cell{2-14}{1}={}{font=\fontsize{0.8em}{1.1em}\selectfont, halign=l},
cell{2-14}{2}={}{font=\fontsize{0.8em}{1.1em}\selectfont, halign=c},
cell{2-14}{5}={}{font=\fontsize{0.8em}{1.1em}\selectfont, halign=c},
}                     
& Log residential rev. &  &  & Log C\&I rev. &  &  \\
& Between & Within & Spills & Between  & Within  & Spills  \\
Log C\&I sales & -0.205*** & -0.101** & -0.073* & 1.496*** & 0.891*** & 0.870*** \\
& (0.014) & (0.035) & (0.035) & (0.027) & (0.060) & (0.061) \\
Log residential sales & 1.158*** & 0.779*** & 0.819*** & -0.627*** & -0.219** & -0.230** \\
& (0.016) & (0.048) & (0.047) & (0.032) & (0.081) & (0.082) \\
Log C\&I sales of neighbors &  &  & -0.020 &  &  & 0.063* \\
&  &  & (0.016) &  &  & (0.028) \\
Log residential sales of neighbors &  &  & -0.081*** &  &  & -0.030 \\
&  &  & (0.020) &  &  & (0.036) \\
Year fixed eff. & X & X & X & X & X & X \\
State fixed eff. &  & X & X &  & X & X \\
$R^2$ & 0.965 & 0.998 & 0.998 & 0.925 & 0.996 & 0.996 \\
Observations & 480 & 480 & 480 & 480 & 480 & 480 \\
\end{talltblr}
}
\end{table}

We now investigate the relationship between sectoral demand and own- and cross-customer-class prices. A natural concern is that data centers, classified as either industrial or commercial customers depending on the jurisdiction, might lower overall system costs while shifting costs to residential ratepayers. Industrial and commercial customers typically see lower rates compared to residential households, so it is plausible that residential customers are subsidizing the incremental fixed costs of these lower-rate customers. To address this concern, we estimate these cross-class effects on residential prices using state-level EIA Form 861 data. Table \ref{tab:demand_rev_corr} shows the estimated correlation between log demand by customer-class and customer-class revenue. The first three columns show results for log residential revenue as the dependent variable, while the last three show log industrial revenue as the dependent variable. Columns 1 and 4 make comparisons across states using year fixed effects, while columns 2 and 5 compare changes within states by using both year and state fixed effects. Columns 3 and 6 include aggregated demand from neighboring states. Note again that coefficients less than 1 indicate a concave correlation and therefore an average price decreasing relationship. 

Comparisons between states show that the correlation of log own-customer-class demand and log revenue is above unity implying that states with higher own-class demand have higher average prices. When we look at own-class demand shifts within states, however,  the relationship is concave implying that an increase in own-class demand is associated with an average price decrease. Additionally, both between and within states, C\&I and residential demand have a negative relationship with the revenue of the other. These estimates both fit the pattern of economies of scale and a lack of systematic cost shifting between customer classes. C\&I demand from neighboring states is associated with decreasing residential and increasing C\&I prices and estimates are statistically significant in both cases at the 0.1\% level. We do not have a clear story for interpreting this difference.

\section{Estimating the Effect of Data Centers on Average Retail Prices}\label{sec:causal}

\begin{figure}[t!]
\begin{subfigure}[b]{0.49\textwidth}
    \caption{Residential -- \%}
    \includegraphics[width=1\textwidth]{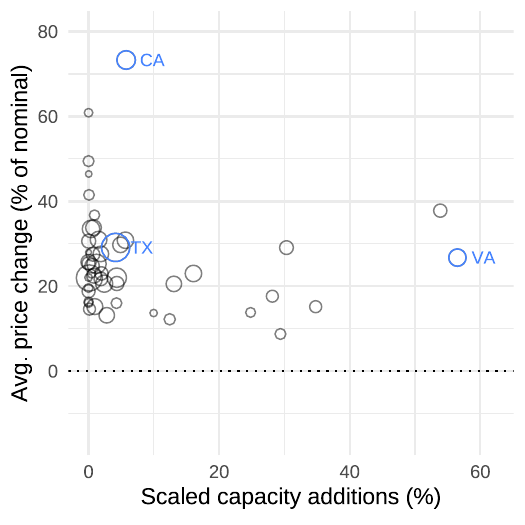}
\end{subfigure}
\begin{subfigure}[b]{0.49\textwidth}
\caption{Commercial \& Industrial -- \%}
\includegraphics[width=1\textwidth]{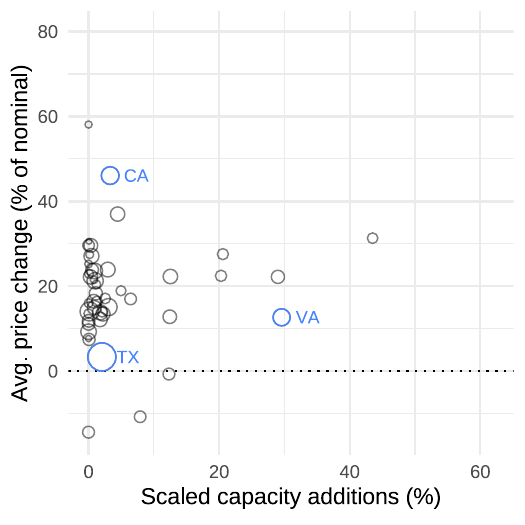}
\end{subfigure}
\begin{subfigure}[b]{1\textwidth}
    \caption{Residential -- levels}
    \includegraphics[width=1\textwidth]{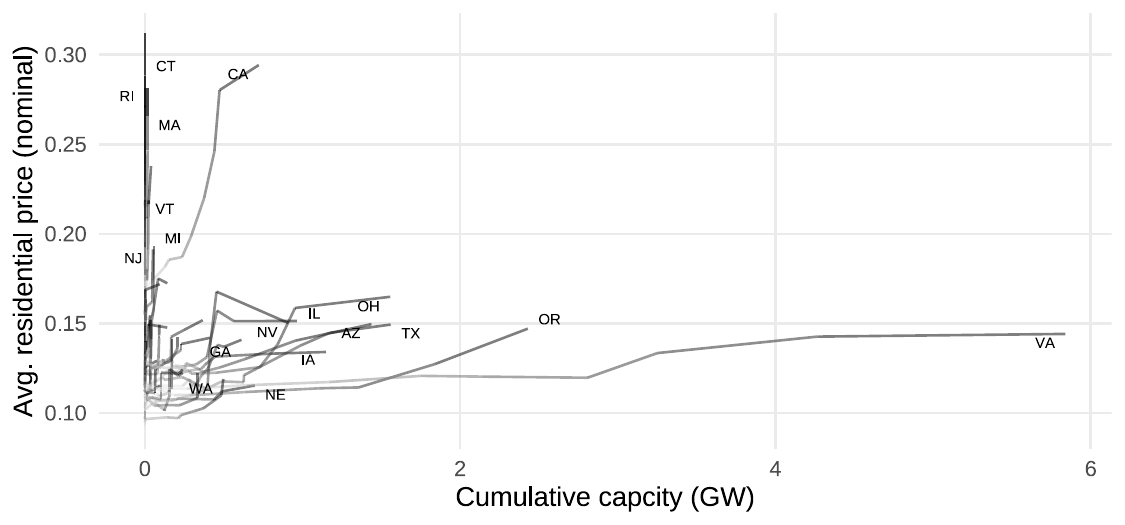}
\end{subfigure}
\caption{Average Price vs Data Center Capacity Change} \label{fig:price_vs_dc}
\begin{figurenotes}

The horizontal axis shows the approximate annual data center demand as a percent of annual sectoral demand from data center capacity additions in 2024 (see text); the vertical axis shows the percent change in sectoral average retail prices in 2024 relative to 2015; circle size is scaled by 2024 sectoral demand. For the bottom panel, the horizontal axis is the cumulative data center capacity in gigawatts; the vertical axis shows the average residential sector price in nominal dollars; lines show the historic change for each state with lighter line segments tracing changes from earlier years. Prices are from EIA Form 861 and data center capacities are from S\&P Global.
\end{figurenotes}
\end{figure}

We now turn to the specific effects of data centers. Figure \ref{fig:price_vs_dc} shows data for scaled data center additions\footnote{As we do not observe power utilization from data centers, we assume a 50\% capacity utilization rate in order to compare capacity additions to sectoral demand. This assumption may be conservative depending on the time it takes to ramp up to maximum annual utilization \citep{epri2026poweringintelligence}.} compared to percent changes in sectoral average price as well as these relationships in levels. Panel (a) shows percent changes for residential customers; Panel (b) shows percent changes for C\&I customers; and Panel (c) shows levels over time. 

Of note is that the state with the highest proportionate residential price change (California) experienced low data center growth relative to sectoral demand, while the state with the most data center growth (Virginia) saw residential price changes close to the average. Data centers and average price changes in percentage terms are negatively correlated. However, it is clear that states that maintained relatively lower average prices saw more data center capacity additions compared to states with large nominal price increases.

To estimate the price effect of data centers, we estimate percent changes in residential revenue conditional on residential demand. We would ideally like to estimate the percent change in total electricity bills holding each customer's demand fixed. Importantly, we do not know the rate structure for any given (or group of) rate payer(s) nor do we observe individual level demand. As percent changes in average all-in prices holding residential demand fixed is exactly equal to the percent change in residential revenue holding residential demand fixed, the effect we estimate is as close as possible to our ideal. Therefore, we estimate the average all-in residential price effect for fixed aggregate residential demand as an elasticity using the functional form:
\begin{equation}\label{eq:reg}
    \ln Rev^{\text{res}}_{st} =  \beta_1 \ln TWh^{\text{res}}_{st} + \beta_2 \sinh^{-1} DC_{st} + \mathbf{\gamma}'X_{st} + \alpha_t + \delta_s  + \varepsilon_{st}
\end{equation}
where: $Rev^{\text{res}}_{st}$ is revenue from  residential consumers in state $s$ and year $t$; $TWh^{\text{res}}_{st}$ is residential sales; $DC_{st}$ is data center capacity additions since 2018 while $\sinh^{-1}$ is the inverse hyperbolic sine operator\footnote{The inverse hyperbolic sine behaves similarly to the natural log, allowing us to estimate elasticities, but, unlike the natural log, allows for zeros in the data, which is necessary in our case. The scale of the variable matters for estimating elasticities \citep{bellemare2020elasticities}. If the mean of the variable is too small, it can cause bias in the interpretation of the estimate. In our case, the mean of capacity additions measured in kilowatts is far above the threshold where bias is rounding error.}; $\beta_2$ is our coefficient of interest and is interpreted as the cross-price elasticity of data centers; $X_{st}$ is a vector of state controls; $\alpha_t$ is a year effect that is constant across all states; $\delta_s$ are state-specific effect that is constant across all time periods. Data centers are either classified as industrial or commercial customers, depending on the state, which may also change over time. Therefore, commercial and industrial sales are omitted as controls as their coefficients would absorb the effect of data centers. In addition to residential price effects we estimate the effects for commercial and industrial customers in the Supplemental Appendix.\footnote{Estimating C\&I demand requires that we switch to log of average industrial price on the left hand side as we cannot control for C\&I demand.}

\subsection{Identification}

We estimate regression \ref{eq:reg} using ordinary least squares (OLS) and two-stage least squares (2SLS) using an instrumental variable (IV). OLS estimates are potentially biased if uncontrolled-for drivers of utility sales revenue correlate with data center capacity. A particular concern is that data center siting decisions may be correlated with expectations about future electricity prices. Indeed, any profit-maximizing investor ought to consider these costs during site selection, which would bias estimates downwards. It is also possible that ``speed to power'' (the length of time for new facilities to be connected to the grid) may be correlated with higher costs, which would bias estimate upwards. The trade press consistently identify ``speed-to-power'' as the dominant factor in siting decisions. 

To the extent that these siting incentives are constant through our study period, the state fixed effects absorb their influence. Data centers could still be endogenous to average prices if the incentives to locate in one state over another are shifting over time. We expect this to be the case for two reasons: (1) the increased pace of data center additions has put new strain on the interconnection queue which would reveal or make relevant differences in utility interconnection ability; (2) recent investments to harden the grid for the warmer climate (e.g. wildfire hardening in California)  could increase expected future prices. These developing phenomena are likely to shift price and ``speed-to-power'' expectations over the study period.

\begin{figure}[h]
\includegraphics[width=1\textwidth]{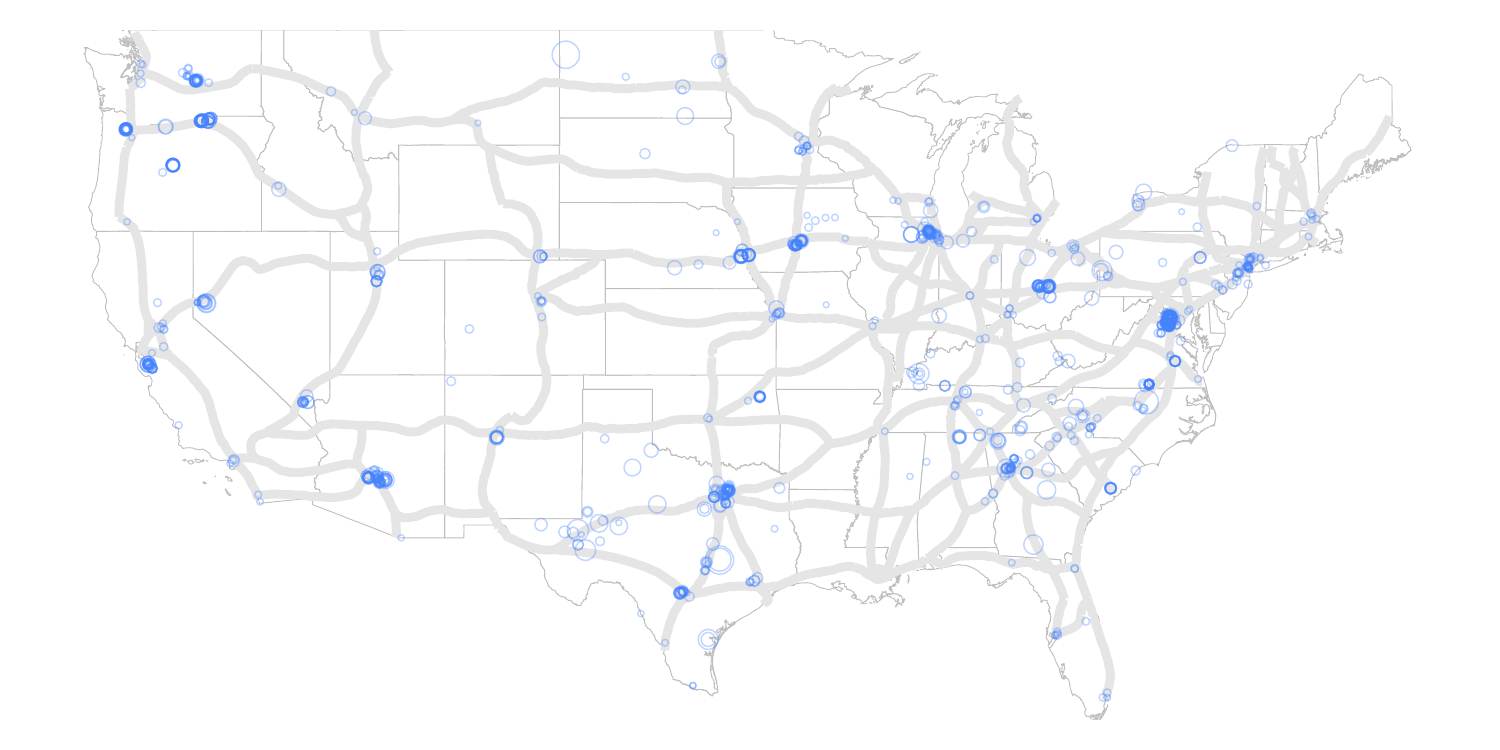}
\caption{Data Center Additions and the 1947 Interstate Highway Plan } \label{fig:iv}
\begin{figurenotes} 
Circles indicate data centers built between 2015 to 2024. The size of the circles is scaled to data center capacity. The thick gray lines show the original 1947 Dwight D. Eisenhower National System of Interstate and Defense Highways Plan. 
\end{figurenotes}
\end{figure}

To overcome these potential sources of endogeneity, we use the total length of the original interstate highway plan as an instrumental variable for data centers. We argue that the 1947 plan affects contemporary electricity prices only through its long-run effects on population, economic activity, and the location of communications infrastructure. By conditioning on state population and GDP, we block the first two channels, thus satisfying the exclusion restriction for instrumental variables. The remaining channel (fiber-optic infrastructure that drives data center siting) is the relevance pathway we exploit. The relevancy of the instrument comes from the observation that terrestrial fiber-optic infrastructure is typically located along interstate highway corridors. This is clearly visible in Figure \ref{fig:iv}, which shows the original 1947 Dwight D. Eisenhower National System of Interstate and Defense Highways Plan overlain with the location of new or expanded data centers during the study period. Further, our instrument is positively correlated with data center capacity additions conditional on the controls (p < 0.001). A regression table for the first stage is provided in the Supplemental Appendix. 

Using IV, the first stage is then:
\begin{equation}
     \sinh^{-1} DC_{st} = \lambda_1 \ln TWh^{\text{res}}_{st} + \lambda_2 z_s + \mathbf{\theta}'X + \zeta_t + \eta_r +\epsilon_{st}.
\end{equation}
For the second stage, equation \ref{eq:reg} becomes:
\begin{equation}
    \ln Rev^{\text{res}}_{st} =  \beta_1 \ln TWh^{\text{res}}_{st} + \beta_2 \reallywidehat{\sinh^{-1} DC_{st}} + \mathbf{\gamma}'X_{st} + \alpha_t + \delta_r +
    \varepsilon_{st}
\end{equation}
where: $z_s$ is the instrument and $\reallywidehat{\sinh^{-1} DC_{st}}$ is the estimated quantity from the first stage. Note that state fixed effects $\delta_s$ are now omitted from the second stage, but regional effects $\delta_r$ are included. This is because the instrument is time-invariant and therefore is collinear with state fixed effects. This creates a tradeoff. By using the instrument, we must rely more heavily on our controls. 

\subsection{Results}

Table \ref{tab:main_results} presents our main OLS and IV estimation results for average residential prices. The first three columns vary by the fixed effects. ``Between'' indicates only year fixed effects were used while ``Within'' adds state fixed effects and ``Within region'' swaps state with regional fixed effects.\footnote{Region definitions are collections of states taken from the 2025 U.S. Regional Economy, Greenhouse Gas, and Energy Model (US-REGEN) energy-economy quantitative equilibrium model developed and maintained by EPRI. The 16 regions in US-REGEN are Pacific: WA, OR; California: CA; Mountain-S: NV, UT, NM, AZ; Mountain-N: ID, MT, CO, WY; MISO-N: ND, MN, IA, WI; SPP: SD, NE, KS, OK; MISO-S: LA, AR, MO; Southeast: MS, AL, GA, TN; Florida: FL; South Atlantic: SC, NC, VA; Ohio Valley: OH, KY, WV; Mid Atlantic: PA, NJ, MD, DE, DC; New York: NY; North East: ME, NH, VT, MA, CT, RI; Texas: TX; MISO-E: IL, MI, IN. These regions roughly correspond to North American Electric Reliability Corp (NERC) Reliability Assessment Areas.} The next two columns are estimated with 2SLS. ``2019-24'' reduces the sample years to those years. All standard errors are robust to heteroskedasticity and are clustered by state to account for serial correlation. Weak instrumental variables are known to cause size inflation. The first-stage F statistic is less than 10 (a commonly used though imperfect critical value for weak instruments) in the first IV estimate, though substantially larger than 10 for ``2019-24''. We compute Anderson-Rubin \citep{anderson1949estimation} p-values and 95\% confidence intervals, which are robust to weak-instruments. 

These results suggest that data centers did not increase residential rates on average. The OLS estimates are all negative and small. Point estimates for within state and within region are nearly identical (-0.003 and -0.001 respectively). This suggests that the regional fixed effects are doing an adequate job at controlling for the selection on states with lower price levels. If developers only sited data centers based on static attributes of states, state fixed-effects would be sufficient for estimating a causal effect, and we would conclude that we could not reject no effect.

\FloatBarrier
\begin{table}
\caption{Data Center Average Residential Price Effects}
\label{tab:main_results}
\centering
\resizebox{1\linewidth}{!}{
\begin{talltblr}[         
caption = {},
  entry = none,
  label = none,
note{}={+ p < 0.1, * p < 0.05, ** p < 0.01, *** p < 0.001},
note{ }={Note: The dependent variable for all columns is log residential revenue. 
An observation is a state-year. The first three columns vary by the fixed effects. ``Between'' 
indicates only year fixed effects were used while ``Within'' adds state fixed effects and ``Within region'' 
swaps state with regional fixed effects. The next two columns are estimated with two-stage-least squares. 
``2019-24'' reduces the sample years to those years. 
The coefficient of interest is $\sinh^{-1}$ data center cap, 
which is the inverse hyperbolic sine of cumulative data center capacity since 2015, 
except in the fifth column, ``2019-24'', where the base year is 2019.
As all regressions are conditional on log residential sales, the effects are 
interpreted as fractional changes in average residential price. 
All standard errors are robust to heteroskedasticity and clustered by state.
The instrumental variable regressions include the Anderson-Rubin 
95\% confidence interval and p-value, which is fully robust to weak instruments.},
]                     
{                     
colspec={Q[]Q[]Q[]Q[]Q[]Q[]},
hline{2}={3,6}{solid, black, 0.03em},
hline{2}={2,5}{solid, black, 0.03em, l=-0.5},
hline{2}={4}{solid, black, 0.03em, r=-0.5},
hline{3}={1-6}{solid, black, 0.05em},
hline{11}={1-6}{solid, black, 0.05em},
hline{1}={1-6}{solid, black, 0.08em},
hline{19}={1-6}{solid, black, 0.08em},
column{3-4,6}={}{font=\fontsize{0.8em}{1.1em}\selectfont, halign=c},
cell{1}{1}={}{font=\fontsize{0.8em}{1.1em}\selectfont, halign=c},
cell{1}{2}={c=3}{font=\fontsize{0.8em}{1.1em}\selectfont, halign=c},
cell{1}{5}={c=2}{font=\fontsize{0.8em}{1.1em}\selectfont, halign=c},
cell{2-18}{1}={}{font=\fontsize{0.8em}{1.1em}\selectfont, halign=l},
cell{2-18}{2}={}{font=\fontsize{0.8em}{1.1em}\selectfont, halign=c},
cell{2-18}{5}={}{font=\fontsize{0.8em}{1.1em}\selectfont, halign=c},
}                     
& OLS &  &  & IV &  \\
& Between & Within state & Within region & Within region  & 2019-24 \\
sinh$^{-1}$ data center cap. & -0.010** & -0.003* & -0.001 & -0.035* & -0.017** \\
& (0.003) & (0.001) & (0.002) & (0.014) & (0.006) \\
Log residential sales & 0.509*** & 0.752*** & 0.826*** & 0.778*** & 0.796*** \\
& (0.064) & (0.087) & (0.063) & (0.041) & (0.036) \\
Log population & 0.555*** & -0.616+ & 0.306** & 0.339*** & 0.327*** \\
& (0.152) & (0.350) & (0.110) & (0.078) & (0.066) \\
Log GDP & -0.053 & 0.159 & -0.104 & 0.029 & -0.062 \\
& (0.114) & (0.205) & (0.088) & (0.069) & (0.056) \\
Year fixed eff. & X & X & X & X & X \\
State fixed eff. &  & X &  &  &  \\
Region fixed eff. &  &  & X & X & X \\
Weak-iv robust 95\%CI &  &  &  & [-0.121,-0.014] & [-0.034,-0.006] \\
Weak-iv robust p-value &  &  &  & 0.000 & 0.001 \\
First-stage F &  &  &  & 7.869 & 21.642 \\
$R^2$ & 0.981 & 0.998 & 0.993 & 0.983 & 0.991 \\
Observations & 480 & 480 & 480 & 480 & 288 \\
\end{talltblr}
}
\end{table}

\FloatBarrier

The first within-region IV estimate, our preferred specification, is an order of magnitude larger than OLS and significant at the 0.1\% level. The estimate can be interpreted thus: a doubling of data center capacity causes residential retail prices to fall by 3.5\% for a fixed level of residential demand. The corresponding Anderson-Rubin confidence interval spans -12.1\% to -1.4\%. The next column reduces the sample to the years 2019 to 2024. For these years the estimated effect is half as much as for the full sample and statistically different from zero at the 0.1\% level. 

In the Supplemental Appendix we provide robustness checks using the first differences estimator and test for neighboring state spillover effects. First differences estimates the impulse response to data centers. The estimate is precisely zero both for OLS and 2SLS with confidence intervals that do not extend much beyond zero. This suggests that the price decreasing effect of data centers takes time, which may explain the smaller effect when restricting the sample years. 

Because power transmission crosses state boundaries, we may be concerned that the price effects of data centers could spill over to other states and bias results. We know from \cite{hausman2025power,ham2025power} that there are substantial grid constraints such that generators in neighboring states are more likely to be dispatched in response to higher load than generators further away. So, any spillovers from data centers ought to be more concentrated regionally. We sum both the instrument and a set of neighbor state controls for all states touching the observed state. The within-state estimate is precisely 0 (standard error < 0.0005) while the IV estimate is positive, smaller in magnitude than the own-state effect, and not statistically significant. Of note is that the instrument performs poorly in this setting (first-stage F of 0.6).

The Supplemental Appendix presents additional robustness checks in the form of coefficient plots including regressions where California or 2020 are dropped from the sample and specifications. These robustness checks yield similar estimates to our preferred specification. We also include controls for new generation capacity additions by technology (wind, solar, natural gas, and energy storage). The point estimate is similar, suggesting that the price-reducing effect operates partly through channels other than generation mix---most plausibly, economies of scale from improved T\&D capital utilization. The Supplemental Appendix also shows sets of results for commercial and industrial average prices. These results are substantially smaller than the effect for residential customer prices. This suggests that the efficiency gains of the power system have mainly accrued to residential customers.

\section{Discussion}\label{sec:discussion}

This study estimates a modest causal impact of data center load growth on decreasing residential prices. This is supported by descriptive statistics and a theoretic model that indicate economies of scale of power systems. Our findings are not without caveats, which we place into three categories: (1) supply constraints, (2) durability of demand growth, and (3) fuel costs.

Several supply constraints are currently impacting the U.S. electric sector: orders on new on natural gas turbines are backlogged, which is bidding up their price \citep{anderson2025turbine}; persistent supply-chain issues limit the availability of transformers and other electrical equipment; high tariffs on photovoltaic arrays from China have increased the price of solar in the United States \citep{gerarden2025strategic}; and aggressive executive action has stalled federal approvals for wind and solar projects while they are litigated in courts. Each of these constraints make new generation capacity investments more expensive than they would be otherwise and possibly more expensive than incumbent supply in some instances. To the extent that these constraints are temporary, price increases can be avoided entirely by delaying interconnection or allocating the inflated incremental cost of grid expansion to rates paid by data centers.

The forward-looking application of our results relies on demand growth being durable. If anticipated load from data centers materially exceeds realized consumption---a possibility raised by current interconnection queue dynamics---utilities risk overbuilding capacity whose fixed costs would be spread across fewer kWhs, reversing the mechanism we identify. Even if demand growth does materialize, associated capacity investments made in advance could temporarily increase average costs. To address these concerns, states including Kansas, Michigan, and Delaware, have pursued financial commitments such as minimum bills or long-term contracts from new large loads.

Finally, our results exclude national-level effects of data center electricity demand such as natural gas price effects. Although we lack the identifying variation to do this, we expect the effect to have been small. A quick back of the envelope calculation shows indirect data center natural gas consumption (via electric generation) rose to about 2\% of total U.S. consumption in 2024\footnote{About 40\% of natural gas in the US is used for electricity according to the EIA and about 4.5\% of electricity was used for data centers in 2024 \citep{epri2026poweringintelligence}.} and 0.04\% of global natural gas consumption or 0.06\% excluding Russia, China, and Iran.\footnote{In 2023 the US consumed an estimated 22.2\% of global natural gas according the IEA while Russia, China, and Iran together consumed 27.7\%.} The US has been a net exporter of natural gas since 2017, which has co-integrated natural gas prices with international oil and coal markets \citep{stock2024market} making the effective market size much larger and the fuel price effect of US data centers much smaller compared to a domestic shut-in market for natural gas.\footnote{Though the blockade of the Strait of Hormuz in 2026 has caused export capacity constraints to bind and shut in the market again.} 

The implications of this study for electrification are optimistic. Shifting from natural gas and fuel oil heating to heat pumps and from gasoline vehicles to electric vehicles is expected to substantially increase electricity demand \citep{bistline2024uses} while providing cost benefits to consumers \citep{epri2026wallet}. To the extent that our results hold for different end-uses of electricity, electrification need not increase prices and may help to lower them. However, this study has abstracted away from the load shape—the pattern of load demand within a day, week, and year—of hypothetical new demand. Electrifying heating and transportation have their own load shapes, which are likely to be quite different than data centers \citep{Bistline_2021}. While our results are encouraging for the rate impacts of electrification, they leave open the question of whether the system can continue to expand quickly enough to accommodate concurrent growth in data center demand and end-use electrification without triggering the supply constraints discussed above.

\newpage

\bibliographystyle{aea}
\bibliography{bib.bib}

\newpage
\appendix
\section{Supplemental Appendix}

\section{Theory of Power System Costs, Prices, and Demand}\label{sec:theory}

We formalize a model of the effect of electricity demand on power system costs and average retail prices. We provide a simplified representation of the power system. We represent the system planner's problem as a nested constrained optimization problem. The inner problem is that of minimizing the cost of providing power in each time step $t$ (e.g. an hour) with fixed grid capacity subject to the constraint of supply meeting demand:
\begin{align}
    C_t=& \
    min_{q_{i,t}}
    \sum_i c_i(q_{i,t}|\mathbf{p^F_t},\mathbf{Q_{t}})\\
    &\text{s.t. }\sum q_{i,t} \geq D_t \ \ \forall t \nonumber
\end{align}
where $c_i$ is the cost function for generator $i$; $q_{i,t}$ is the quantity of power from generator $i$ at time $t$; $\mathbf{p^F_t}$ is a vector of fuel prices, and $\mathbf{Q_{t}}$ is the abstracted total capacity of the grid inclusive of generators, transmission, and distribution;$D_t$ is the aggregate demand; $C_t$ is the value function---the solution to the cost minimization problem.\footnote{A more detailed model would include additional constraints, such as generator limits, power balance at each node, and transmission limits. See \citealt{kirschen2026fundamentals}.}

In the long-run, the system planner solves the capacity expansion problem, minimizing the present value of total costs by choosing investments in new capacity ($x$) and capacity retirements ($r$): 
\begin{align}\label{eq:planners_problem}
    TC =& \min_{x_t} \sum_{t} \delta_t \Bigg[ I(x_t) + O(Q_t) + C_t(Q_t, D_t) \Bigg]\\
    \text{s.t.}&\ Q_t = Q_{t-1} + x_{t} - r_{t}, \nonumber \\
    &\ Q_t \geq (1+m)D_t\nonumber, \\
    &\ x_t \geq 0. \nonumber
\end{align}
Here $\delta_t$ is the discount factor, $I$ is the investment cost function of expanding grid capacity by $x$;  $O(Q_t)$ are fixed operation and maintenance costs. The first constraint is the equation of motion for total capacity. The second constraint ensures that capacity satisfies the required reserve margin, denoted by $m$. The third constraint enforces non-negativity of new capacity.\footnote{For simplicity, we omit the capacity availability coefficient ($\in [0,1]$) on non-dispatchable energy (wind and solar), policy constraints, and other operational constraints and assume that retirements are exogenous. Policy constraints such as Renewable Portfolio Standards,  and subsidies for investment and production can be thought of as baked into $I$, $C$, and $x$.} Note that this formulation represents a planner that serves all three components of the system (similar to a vertically integrated utility), rather than a retailer of wholesale power (as in a restructured market). The relevant difference between the two is that the restructured utility does not have generation investment or operation costs, but does build and maintain the transmission and distribution network, and pays a combination of market clearing wholesale prices and contracted rates for electricity. 

Without a binding capacity constraint, new capacity is added only if it reduces short-run operating costs by shifting the dispatch curve outwards. Assume the following convexity conditions: (1) these savings from capacity expansion are declining with capacity, or $\partial^2 C_t / \partial x^{*2}_t>0$; (2) the sum of marginal investment and operation costs are either increasing in capacity expansion or declining at a slower rate than cost savings, or $\partial^2 I / \partial x^{*2}_t + \partial^2 O / \partial x^{*2}_t > - \sum_t \delta_t \partial^2 C_t / \partial x^{*2}_t$; and, (3) capacity investments last for the entire planning horizon. Under these conditions, the solution to the capacity expansion problem is characterized by the necessary condition: 
\begin{equation}\label{eq:FOC}
     \frac{\partial I}{\partial x^*_t} + \frac{\partial O}{\partial x^*_t} = 
     - \sum_t \delta_t \frac{\partial C_t}{\partial x^*_t} + \lambda^*_t
\end{equation}
where $\lambda$ is the shadow cost of the capacity constraint (possibly slack). That is, the planner will invest in capacity up to the point where the marginal cost of new capacity equals the marginal value of new capacity plus the future stream of marginal savings from capacity expansion in terms of reduced variable costs. 

Define $D_T$ as a demand shock that is constant and lasts $T$ periods. Using the envelope theorem,\footnote{$\mathcal{L^*} = \sum_t \delta_t \big(I(x_t^*) + O(x_t^*) + C_t(x_t^*, D_t|Q_{t-1},r) - \lambda^*_t \left[ Q_{t-1} + x_{t}^* - r_{t} - (1+m)D_t\right]\big)$} the change in total cost from a durable demand shock is given by the expression: 
\begin{align}\label{eq:result}
    \frac{\partial TC^*}{\partial D_T}= 
    \frac{\partial \mathcal{L}^*}{\partial D_T}= 
    \sum_t^T \delta_t
     \frac{\partial C_t}{\partial D_t} + \lambda_t^*(1+m).
\end{align}
The change in total costs is equal to the discounted sum of short-run marginal cost plus the shadow cost of the capacity constraint, which is zero if the constraint is slack. With a non-binding capacity constraint, demand can grow without adding new capacity (although capacity expansion may be cost-effective). When the capacity constraint is binding, new capacity must be added in response to an increase in demand. By re-arranging Equation \ref{eq:FOC} and plugging into Equation \ref{eq:result}, we can solve for the total cost effect of new demand in these two regimes:
\begin{align}
         \frac{\partial TC^*}{\partial D_T}
    &= 
    \sum_t^T \delta_t
    \frac{\partial C_t}{\partial D_t}
        & \textit{(slack)} \label{eq:regeime1}\\
    \frac{\partial TC^*}{\partial D_T}
    &= 
      \sum_t^T \delta_t
    \frac{\partial C_t}{\partial D_t} 
    + (1 + m) \Bigg(
    \frac{\partial I}{\partial x^*_t} + \frac{\partial O}{\partial x^*_t}
    + \sum_t^T \delta_t
     \frac{\partial C_t}{\partial x^*_t}
    \Bigg)
          & \textit{(binding)}.\label{eq:regeime2}
\end{align}
In the first regime, the change in total cost is simply equal to the discounted future stream of marginal variable costs. Since the system was already at an optimum,\footnote{The system is at an optimum in the sense that new capacity is added until the marginal costs equals marginal benefit, not necessarily in the sense that all assets on the grid would be optimal if built anew with the technologies and policies of today. With changes in technology and policy and long-lived capital stock, existing power systems certainly do not satisfy the second sense, and may not satisfy the first either due to various market and information barriers.} the marginal cost of increasing capacity and the marginal savings from that increased capacity exactly offset. In the second regime, the change in total cost is again equal to the the stream of marginal variable cost (first term) plus the incremental cost of new capacity net of the variable cost savings from shifting the demand curve scaled by the reserve margin (second term), which is positive in the binding case.\footnote{This result holds for a restructured utility as well given the same set of assumptions, which in this case implies a competitive market for capacity and generation with free entry. In this case, the incentives for capacity expansion remain the same. With a demand shock the cost of power and capacity expansion is equal to the status quo at the margin, which gives the result in Equations \ref{eq:regeime1} and \ref{eq:regeime2}. For large demand shocks or where capacity expansion is constrained there is an additional cost increase compared to the vertically integrated utility case. If the demand shock is large enough to increase market clearing prices for electricity and/or capacity, then the cost of generation increases for all inframarginal units that are not bound by contracted rates. In the extreme case where there are no long-term generation contracts, the total costs increase both from the cost of building and dispatching incremental generation ($\Delta D\cdot(p_{t-1}+\lambda_{t-1})$) and from rents accruing to  inframarginal units ($D_t\cdot(\Delta \lambda + \Delta p)$).}

Importantly, this result holds for small changes in demand, when there is an interior solution, and when capacity expansion is not constrained. If the demand shock takes the system far from the initial optimum, change in total cost will include the shift along the dispatch curve and, if there is no capacity expansion constraint, the cost of the induced capacity expansion net of the savings from the outward shift of the dispatch curve from this capacity. An expression for this type of large shift in demand is given in the subsection below.

In retail markets for electricity, neither average retail prices nor volumetric ``energy-only'' prices coincide with the marginal cost of incremental supply \citep{borenstein2022two}. Instead, customers are charged varying combinations of fixed and volumetric rates, which are expected to recover the utility's costs and allow for some profit, depending on the utility structure \citep{wilson1993nonlinear}. Utilities may also offer tiered pricing, time of use pricing, or special pricing for specific uses such as electric vehicle charging. So, it is common practice to use average all-in prices to make prices comparable between utilities (e.g. \citealt{dong_etal_2025,borenstein2022two,holland2016there}). Average prices $\bar{p}_t$ are simply total revenue $TR_t$ divided by total sales $D_t$:
\begin{equation}
    \bar{p}_t =TR_t/D_t,
\end{equation}
For this exercise, we assume that rates are set such that total revenue equals total costs in the current time step ($TR_t = \dot{TC}$). Employing the classic result that average costs are increasing in quantity if and only if marginal costs are greater than average costs, the conditions for a price decreasing effect are
\begin{align}\label{eq:sign}
    \bar{p}_t
    &> \frac{\partial C_t}{\partial D_t} 
    &\textit{(slack)} \\
     \bar{p}_t
    &> \frac{\partial C_t}{\partial D_t} + 
    \frac{1+m}{T}\Big(
    \frac{\partial I}{\partial x^*_t} + \frac{\partial O}{\partial x^*_t}
    + \sum_t^T \delta_t
     \frac{\partial C_t}{\partial x^*_t}
    \Big)
    &\textit{(binding)}.
\end{align}

In the first regime, where the capacity constraint is non-binding, an increase in demand decreases average costs as long as average rates exceed the short-run marginal cost. Because of the large share of fixed cost included in average rates, this condition holds nearly everywhere. For example, the average wholesale power price in the Northeast was around 4 cents per kWh, compared to an average retail rate in the region above 10 cents and a national sales-weighted average of 12 cents for the period 2014 to 2016 (\cite{borenstein2022two}). With marginal costs approximately half of average costs, a 10\% increase in demand should cause average prices to fall by 5\% when there is excess capacity. 

In the second regime, the incremental cost of supply also includes the levelized fixed cost of a marginal capacity increase net of the variable cost savings from capacity, scaled by the reserve margin. Even including this second term, the price effect remains negative as long as the net cost of new amortized cost of new capacity is lower than the embedded fixed costs in current average rates. For example, combined cycle natural gas (CCNG) typically has a levelized fixed cost of about \$0.02 per kWh \citep{epri2024generation}. The recent CCNG supply shortage has increased capital costs by about 60\% \citep{epri2026tag}. Assuming CCNG represents the marginal technology for capacity expansion and a 15\% reserve margin, the second term is less than \$0.04 per kWh, which is comparable to recent highs in the PJM capacity market.\footnote{Although new CCNG is a prevalent option for meeting new demand, simple cycle natural gas fired combustion turbines are likely the price-setting technology in recent capacity auctions in PJM and elsewhere.} Thus, the sum of short-run marginal cost term and the new capacity term remain well below current average rates across CONUS. Even if some other combination of technologies is on the margin, we expect the incremental cost of capacity to be lower than incumbent costs due to long-run secular decreases in fixed costs and efficiency improvements. 

This exercise elucidates the importance of the ``build margin'' when estimating the effect of long-lived demand on the power system. Existing studies that exploit short-run random variation \citep{holland2016there,holland2019distributional,dong_etal_2025} are only able to capture the dispatch effect (``operating margin''). While the operating margin is informative, as Equation \ref{eq:result} shows, these studies ignore the build margin, which is introduced with the binding capacity constraint, as Equation \ref{eq:regeime2} shows.

Essentially, whether incremental costs are lower than average prices is equivalent to whether total cost with respect to demand is concave or convex. In the subsection below we prove the lemma that if $TC^*(D_T)$ is concave, then $\partial p_t / \partial D_T$ is negative. Because of the high share of fixed costs in the production and delivery of electricity, and because of technical change over time amid long-lived capital assets, we expect total costs to be concave, i.e. to exhibit increasing returns to scale. However, various market conditions potentially confound this result. We test this hypothesis empirically using both utility cost data and state-level data for retail prices and sales.

\subsection{\textbf{Lemma} if $f(q)$ is concave, then $\frac{\partial f(q)/q}{\partial q}$ is negative.}

Proof sketch for the case $f(q)$ strictly concave. Assume $q \geq 0$, and $f(0)\geq0$ (costs are never negative). Let $g(q)=f(q)/q$. We want to show $f(q)$ strictly concave implies $g'(q)<0$. $g'(q)=[f'(q)q-f(q)]/q^2$ implies $\text{sign}(g'(q))=\text{sign}(f'(q)q - f(q))$. $f(q)$ strictly concave implies $f(y)<f(x) + f'(x)(y-x)$. Therefore, $0 \geq f(0) < f(x)-f'(x)\cdot x$. So, $0 \leq f(0)\leq f(q)-f'(q)x$ and $f'(q)q-f(q)<0$ and $g'(q)<0$.

\subsection{Change in total costs for large demand shifts}
This subsection shows the mathematical expression for how total costs change when the demand shift is large in the case without capacity constraints. We use a second order Taylor-series approximation, which yields:
\begin{multline}
\frac{\Delta TC^*}{\Delta DT}
    \approx 
    \sum_t^T \delta_t \frac{\partial C_t}{\partial D_t}\Delta D \\
    + \frac{1}{2}\Bigg\{
     \underbrace{\sum_t^T \delta_t \left[\frac{\partial^2 C_t}{\partial D_t^2} + \frac{\partial^2 C_t}{\partial x_t \partial D_t}\frac{\partial x^2_t}{\partial D_T}\right]}_{\text{curvature of variable costs w.r.t. demand}} \nonumber \\
    +\underbrace{\left[\frac{\partial^2 I}{\partial x^{2}_t} 
    + \frac{\partial^2 O}{\partial x^2_t} 
    + \sum_t^T \delta_t \frac{\partial^2 C_t}{\partial x^2_t}\right]\left(\frac{\partial x_t}{\partial D_T}\right)^2}_{\text{net curvature of capacity re-optimization}} 
    \Bigg\}(\Delta D)^2 
    \nonumber
\end{multline}
Here we see that total costs expand by the rate of current volumetric costs (first term) as before. But, cost also change due to the net effect of movement along the dispatch curve (second term) and the shift in the dispatch curve from capacity expansion (third term). For small shifts in demand from an equilibrium these last two terms cancel. 

\section{Supplemental Tables and Figures}

\FloatBarrier
\begin{table}
\caption{Sectoral Demand and Cost Relationship}
\label{tab:demand_cost_corr}
\centering
\resizebox{0.75\linewidth}{!}{
\begin{talltblr}[         
entry=none,label=none,
note{}={Note: The dependent variable is the row header and each row
               reports estimates from 5 separate regressions. 
               Columns declare the model specification. Column 1 is the ordinary 
               least squares
               estimate, column 2 adds year fixed effects, and column three includes
               both year and utility fixed effects
               The 95\% confidence interval are shown below the point estimate 
               in brackets. All estimates are less than 1, indicating a negative 
               correlation between with average prices. 
               There are between 414 and 440 observations in each model depending 
                        on missing data.},
]                     
{                     
colspec={Q[]Q[]Q[]Q[]},
hline{2}={1-4}{solid, black, 0.05em},
hline{1}={1-4}{solid, black, 0.1em},
hline{29}={1-4}{solid, black, 0.1em},
column{2-4}={}{halign=c},
cell{1,3-6,8-11,13-16,18-21,23-28}{1}={}{halign=l},
cell{2,7,12,17,22}{1}={c=4}{halign=l},
}                     
& OLS & Between & Within \\
log T\&D opex & log T\&D opex & log T\&D opex & log T\&D opex \\
\ \ \ \ log res. MWh & 0.40 & 0.41 & 0.06 \\
& [0.28, 0.52] & [0.28, 0.53] & [-0.13, 0.24] \\
\ \ \ \ log C\&I MWh & 0.45 & 0.44 & -0.04 \\
& [0.33, 0.57] & [0.32, 0.56] & [-0.16, 0.07] \\
log T\&D capex & log T\&D capex & log T\&D capex & log T\&D capex \\
\ \ \ \ log res. MWh & 0.71 & 0.73 & 0.41 \\
& [0.61, 0.82] & [0.63, 0.83] & [-0.08, 0.89] \\
\ \ \ \ log C\&I MWh & 0.35 & 0.33 & -0.45 \\
& [0.24, 0.45] & [0.23, 0.43] & [-0.76, -0.15] \\
log Gen. opex & log Gen. opex & log Gen. opex & log Gen. opex \\
\ \ \ \ log res. MWh & 0.46 & 0.47 & 0.19 \\
& [0.39, 0.54] & [0.40, 0.54] & [-0.03, 0.42] \\
\ \ \ \ log C\&I MWh & 0.48 & 0.46 & 0.25 \\
& [0.40, 0.55] & [0.39, 0.53] & [0.11, 0.40] \\
log Gen capex & log Gen capex & log Gen capex & log Gen capex \\
\ \ \ \ log res. MWh & 0.52 & 0.54 & 0.58 \\
& [0.26, 0.78] & [0.28, 0.80] & [-0.79, 1.95] \\
\ \ \ \ log C\&I MWh & 0.77 & 0.75 & -0.07 \\
& [0.50, 1.04] & [0.48, 1.02] & [-0.94, 0.81] \\
log Total cost & log Total cost & log Total cost & log Total cost \\
\ \ \ \ log res. MWh & 0.52 & 0.36 & 0.64 \\
& [0.26, 0.78] & [0.21, 0.51] & [0.09, 1.20] \\
\ \ \ \ log C\&I MWh & 0.77 & 0.68 & -0.53 \\
& [0.50, 1.04] & [0.53, 0.82] & [-0.88, -0.18] \\
\midrule
Year fixed eff. &  & X & X \\
Util. fixed eff. &  &  & X \\

\end{talltblr}
}
\end{table}

\FloatBarrier

\begin{table}
\caption{First-stage Regression Results}
\label{tab:first_stage}
\centering
\begin{talltblr}[         
caption = {},
  entry = none,
  label = none,
note{}={+ p < 0.1, * p < 0.05, ** p < 0.01, *** p < 0.001},
]                     
{                     
colspec={Q[]Q[]Q[]Q[]Q[]},
hline{2}={1-5}{solid, black, 0.05em},
hline{18}={1-5}{solid, black, 0.05em},
hline{1}={1-5}{solid, black, 0.08em},
hline{22}={1-5}{solid, black, 0.08em},
column{2-5}={}{font=\fontsize{0.8em}{1.1em}\selectfont, halign=c},
column{1}={}{font=\fontsize{0.8em}{1.1em}\selectfont, halign=l},
}                     
& Within & 20219-24 & 1st diff & Spills \\
Instrument & 1.297** & 3.023*** & 2.967*** & -0.597 \\
& (0.462) & (0.650) & (0.301) & (0.733) \\
Log residential sales & -0.902 & -1.335 & -3.904 & 4.779 \\
& (0.947) & (1.343) & (5.564) & (3.396) \\
Log C\&I sales &  &  &  & 0.991 \\
&  &  &  & (1.720) \\
Log population & -0.218 & -3.792 & 39.390 & -13.730** \\
& (1.622) & (2.335) & (38.688) & (5.038) \\
Log GDP & 4.403*** & 6.323*** & 2.613 & 10.444** \\
& (1.158) & (1.665) & (11.130) & (3.882) \\
Log residential sales (neighbors) &  &  &  & 5.412*** \\
&  &  &  & (1.072) \\
Log population (neighbors) &  &  &  & -3.211* \\
&  &  &  & (1.525) \\
Log GDP (neighbors) &  &  &  & 2.221*** \\
&  &  &  & (0.622) \\
Year fixed eff. & X & X & X & X \\
Region fixed eff. & X & X & X & X \\
$R^2$ & 0.728 & 0.667 & 0.407 & 0.834 \\
Observations & 480 & 288 & 432 & 480 \\
\end{talltblr}
\end{table}

\begin{table}
\caption{Data Center Average Residential Price Effects, Robustness Checks}
\label{tab:robust_results}
\centering
\resizebox{1\linewidth}{!}{
\begin{talltblr}[         
caption = {},
  entry = none,
  label = none,
note{}={+ p < 0.1, * p < 0.05, ** p < 0.01, *** p < 0.001},
note{ }={Note: The dependent variable for all columns is log residential revenue. 
An observation is a state-year. ``1st diff'' is the first difference estimator. 
The columns with the header ``Spills'' estimate the spillover effect from neighbors. 
All standard errors are robust to heteroskedasticity and clustered by state.
The instrumental variable regressions include the Anderson-Rubin 
95\% confidence interval and p-value, which is fully robust to weak instruments.},
]                     
{                     
colspec={Q[]Q[]Q[]Q[]Q[]},
hline{2}={5}{solid, black, 0.03em},
hline{2}={2,4}{solid, black, 0.03em, l=-0.5},
hline{2}={3}{solid, black, 0.03em, r=-0.5},
hline{3}={1-5}{solid, black, 0.05em},
hline{21}={1-5}{solid, black, 0.05em},
hline{1}={1-5}{solid, black, 0.08em},
hline{28}={1-5}{solid, black, 0.08em},
column{3,5}={}{font=\fontsize{0.8em}{1.1em}\selectfont, halign=c},
cell{1}{1}={}{font=\fontsize{0.8em}{1.1em}\selectfont, halign=c},
cell{1}{2}={c=2}{font=\fontsize{0.8em}{1.1em}\selectfont, halign=c},
cell{1}{4}={c=2}{font=\fontsize{0.8em}{1.1em}\selectfont, halign=c},
cell{2-27}{1}={}{font=\fontsize{0.8em}{1.1em}\selectfont, halign=l},
cell{2-27}{2}={}{font=\fontsize{0.8em}{1.1em}\selectfont, halign=c},
cell{2-27}{4}={}{font=\fontsize{0.8em}{1.1em}\selectfont, halign=c},
}                     
& 1st Diff &  & Spills &  \\
& Within & IV & Within  & IV  \\
sinh$^{-1}$ data center cap. & 0.000 & 0.000 &  &  \\
& (0.000) & (0.001) &  &  \\
sinh$^{-1}$ data center cap. of neighbors &  &  & 0.000 & 0.007 \\
&  &  & (0.000) & (0.012) \\
Log residential sales & 0.849*** & 0.849*** & 0.847*** & 0.749*** \\
& (0.063) & (0.081) & (0.088) & (0.066) \\
Log C\&I sales &  &  & -0.073 & -0.033+ \\
&  &  & (0.048) & (0.019) \\
Log residential sales of neighbors &  &  & -0.051+ & -0.101+ \\
&  &  & (0.027) & (0.061) \\
Log population & -0.627+ & -0.627 & -0.628+ & 0.386* \\
& (0.345) & (0.392) & (0.341) & (0.177) \\
Log GDP & 0.107 & 0.107 & 0.254 & -0.121 \\
& (0.149) & (0.112) & (0.169) & (0.139) \\
Log population of neighbors &  &  & -0.028 & 0.119** \\
&  &  & (0.096) & (0.040) \\
Log GDP of neighbors &  &  & -0.051+ & -0.039 \\
&  &  & (0.027) & (0.026) \\
Year fixed eff. & X & X & X & X \\
Region fixed eff. &  &  & X & X \\
Weak-iv robust 95\%CI &  & [-0.002,0.002] &  & [-Inf,Inf] \\
Weak-iv robust p-value &  & 0.872 &  & 0.439 \\
First-stage F &  & 96.896 &  & 0.663 \\
$R^2$ & 0.533 & 0.533 & 0.998 & 0.991 \\
Observations & 432 & 432 & 480 & 480 \\
\end{talltblr}
}
\end{table}

\begin{figure}
\caption{Residential Revenue Estimates}
\includegraphics[width=1\textwidth]{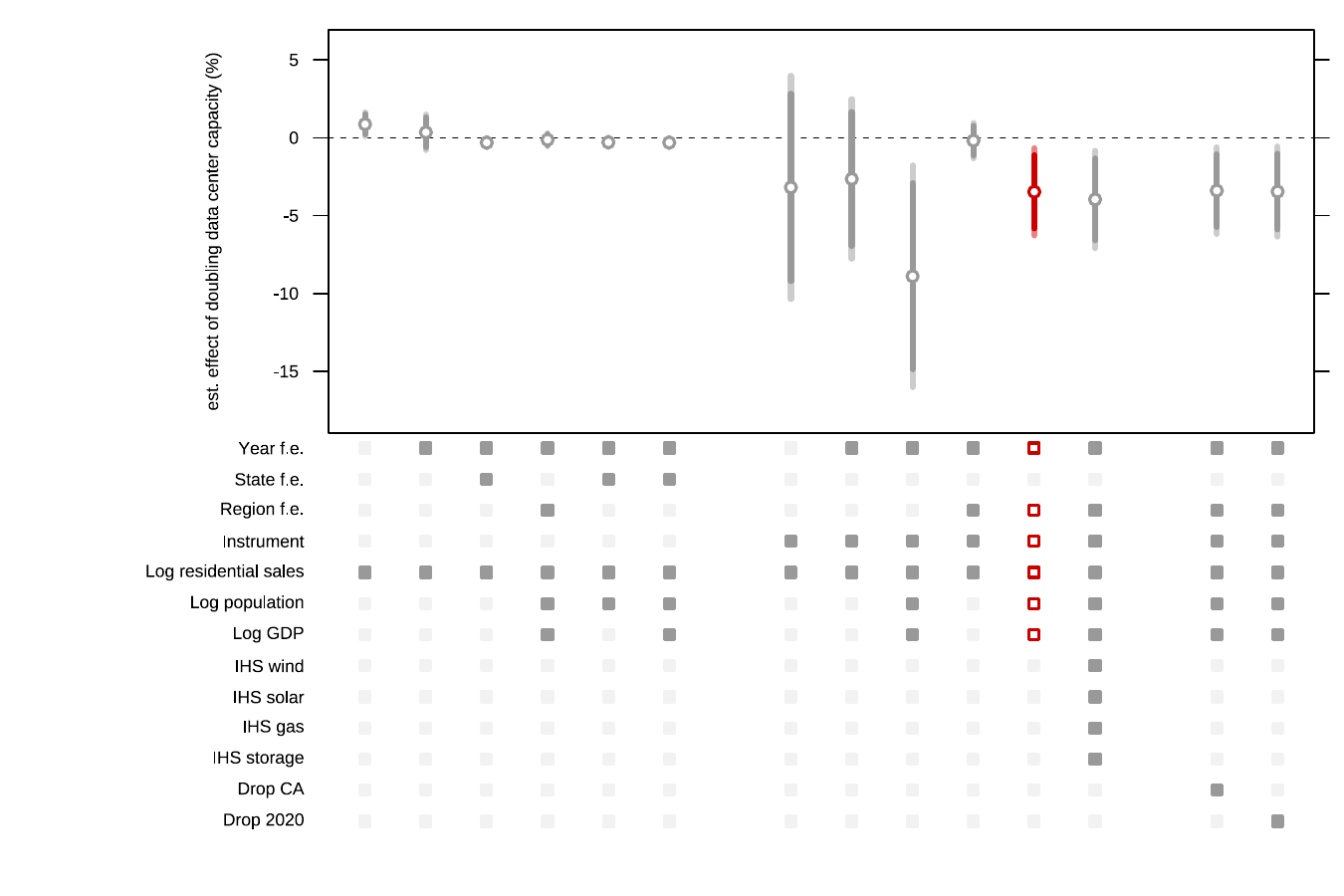}
\begin{figurenotes}
    The dependent variable is the log of average residential revenue. The highlighted column corresponds to column 4 in the main results. The vertical axis shows the scaled coefficient as the percent change in response to a doubling of data center capacity (coefficients are multiplied by 100). 90\% and 95\% confidence intervals shown are clustered by state but are not Anderson-Rubin confidence intervals, which are typically tighter.
\end{figurenotes}
\end{figure}

\begin{figure}
\caption{Industrial Average Price Estimates}
\includegraphics[width=1\textwidth]{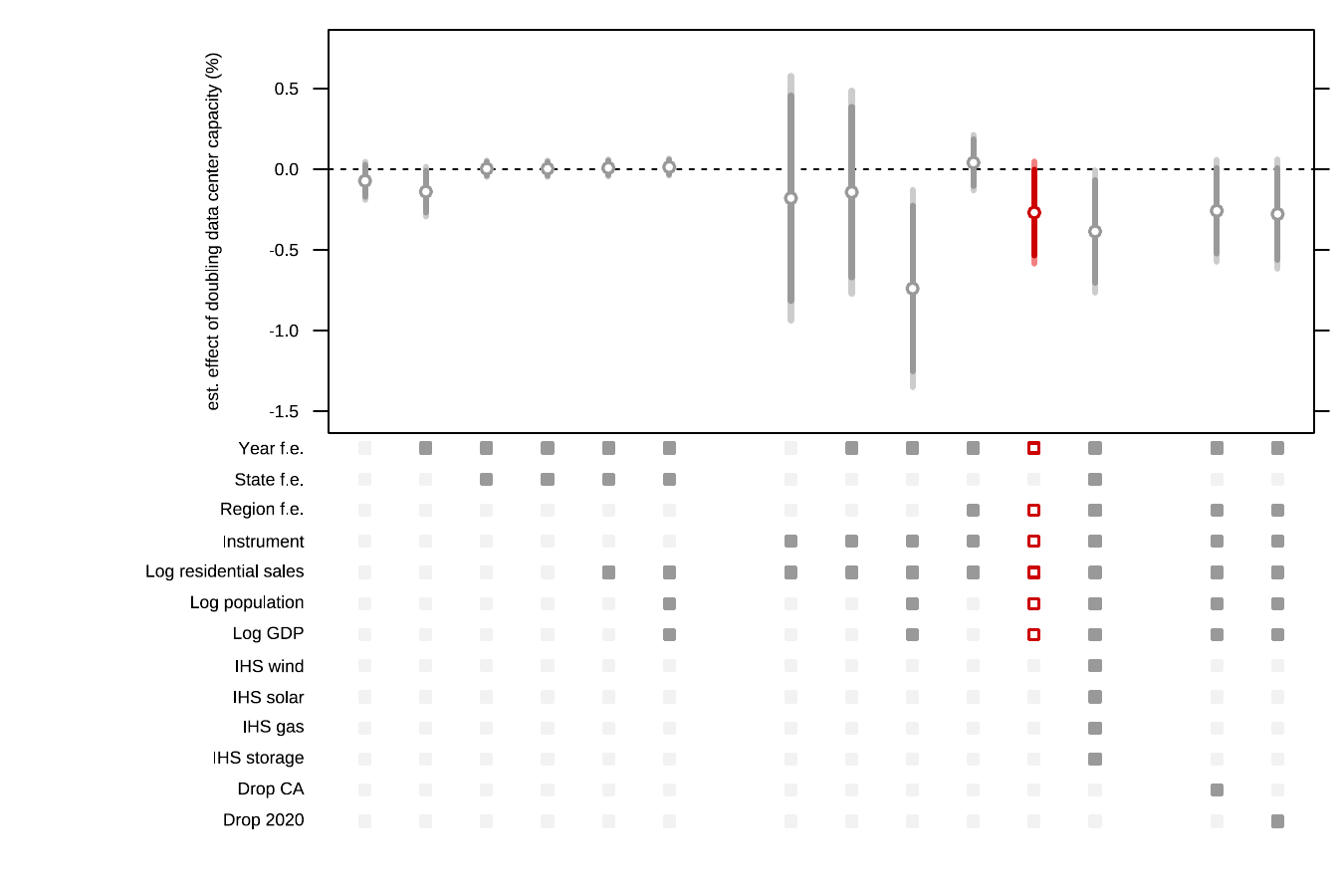}
\begin{figurenotes}
    The dependent variable is the log of average industrial price. The vertical axis shows the scaled coefficient as the percent change in response to a doubling of data center capacity (coefficients are multiplied by 100). 90\% and 95\% confidence intervals shown are clustered by state but are not Anderson-Rubin confidence intervals, which are typically tighter.
\end{figurenotes}
\end{figure}

\begin{figure}
\caption{Commercial Average Price Estimates}
\includegraphics[width=1\textwidth]{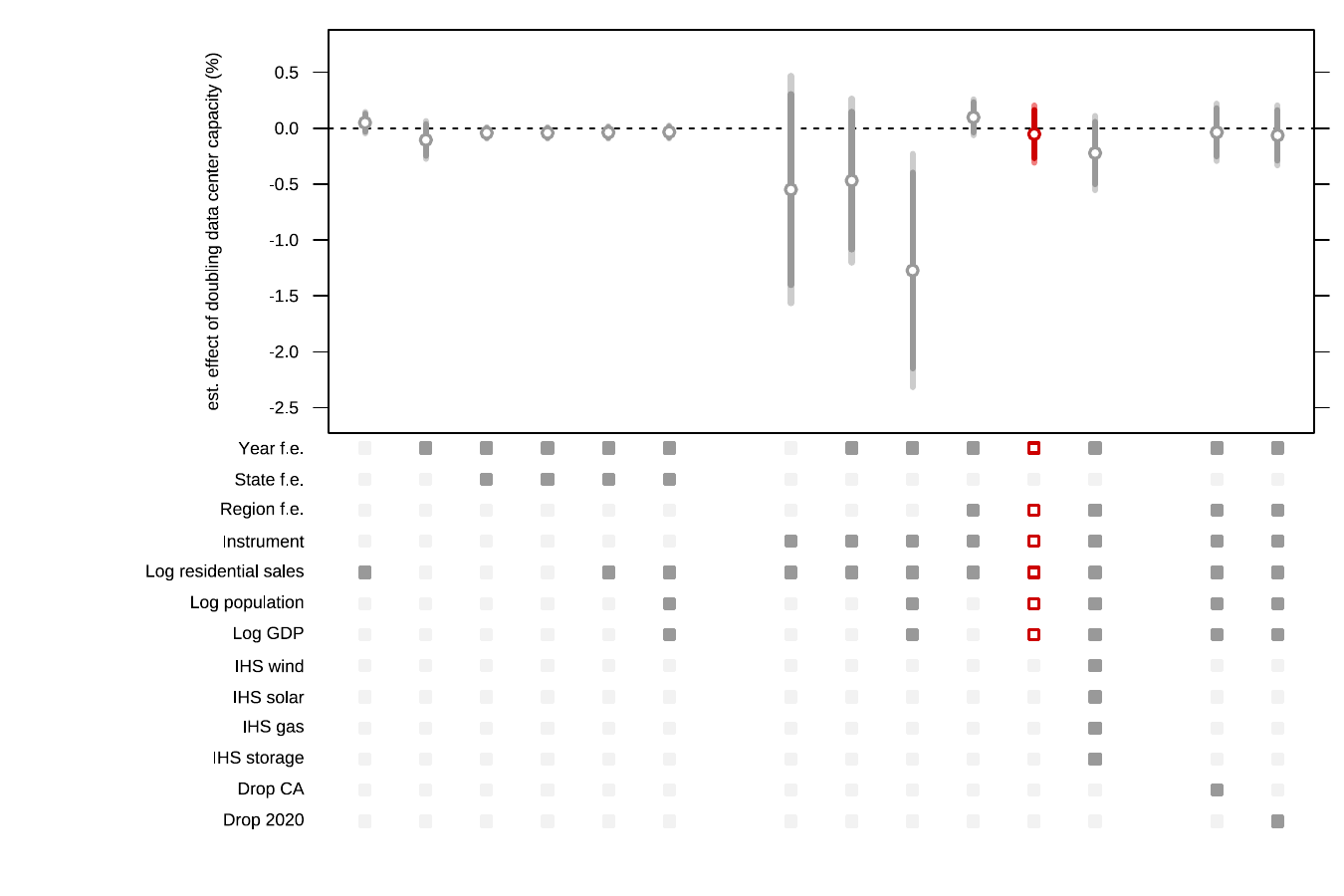}
\begin{figurenotes}
    The dependent variable is the log of average commercial price. The vertical axis shows the scaled coefficient as the percent change in response to a doubling of data center capacity (coefficients are multiplied by 100). 90\% and 95\% confidence intervals shown are clustered by state but are not Anderson-Rubin confidence intervals, which are typically tighter.
\end{figurenotes}
\end{figure}

\end{document}